\documentclass[a4paper,12pt]{article}
\usepackage{graphicx,color, pifont,psfrag}
\usepackage{amsmath,amssymb,a4wide,epsfig,axodraw}
\date{\today}
 \normalsize

\newcommand{\bmat}{\left(\begin{array}}
\newcommand{\emat}{\end{array}\right)}
\newcommand{\be}{\begin{equation}}
\newcommand{\ee}{\end{equation}}
\newcommand{\bea}{\begin{eqnarray}}
\newcommand{\eea}{\end{eqnarray}}

\def\lsim{\raise0.3ex\hbox{$\;<$\kern-0.75em\raise-1.1ex\hbox{$\sim\;$}}}
\def\gsim{\raise0.3ex\hbox{$\;>$\kern-0.75em\raise-1.1ex\hbox{$\sim\;$}}}

\definecolor{grey}{rgb}{0.3,0.3,0.3}
\def\corrE#1{\textcolor{grey}{#1}}

\begin{document}
\bibliographystyle{unsrt}
\renewcommand{\thefootnote}{\fnsymbol{footnote}}
\rightline{LPT-ORSAY-09-14} \rightline{LAL-09-28}
\vspace{.3cm} 
{\Large
\begin{center}
{\bf Non-leptonic charmless $B_c$ decays\\
and their search at LHCb}
\end{center}}
\vspace{.3cm}

\begin{center}
S. Descotes-Genon$^a$, J. He$^b$, E. Kou$^b$ and P. Robbe$^b$\\

\vspace{.3cm}\small
\emph{$^a$ Laboratoire de Physique Th\'eorique, 
CNRS/Univ. Paris-Sud 11 (UMR 8627)}\\
\emph{91405 Orsay, France}

\emph{$^b$ 
Laboratoire de l'Acc\'el\'erateur Lin\'eaire,
Universit\'e Paris-Sud 11, CNRS/IN2P3 (UMR 8607)} \\
\emph{91405 Orsay, France}

\end{center}

\vspace{.3cm}
\hrule \vskip 0.3cm
\begin{center}
\small{\bf Abstract}\\[3mm]
\end{center}
We discuss the decay of $B_c$ mesons into two light mesons ($\pi, K^{(*)},
\eta^{(\prime)}, \rho, \omega, \phi$). All these decay channels come from a
single type of diagram, namely tree annihilation. This allows us to derive
extremely simple $SU(3)$ relations among 
these processes. The size of annihilation contributions is an important issue
in $B$ physics, and we provide two different estimates in the case of
non-leptonic charmless $B_c$ decays, either a comparison with
annihilation decays of heavy-light mesons or 
a perturbative model inspired by QCD factorisation.
We finally discuss a possible search for these channels at LHCb.
\begin{minipage}[h]{14.0cm}
\end{minipage}
\vskip 0.3cm \hrule \vskip 0.8cm
%%%%%%%%%%%%%%%%%%%%%%%%%%%%%%%%%%
%
\section{{\large \bf Introduction}}\label{sec:1}
%%%%%%%%%%%%%%%%%%%%%%%%%%%%%%%%%%%
The investigation of the properties of the $B_c$ meson started in 1998 when
the CDF collaboration observed 20.4 events containing a
$B_c$ in the channel $B_c\to J/\psi l\nu$~\cite{Abe:1998wi}. Since then, its mass and width have been measured~\cite{Aaltonen:2007gv},
and bounds on some non-leptonic channels have been set ($J/\psi$ with
one or three pions, $D^{*+}\bar{D}^0$\ldots). From the theoretical point of view,
the $B_c$ meson shares many features with the better known quarkonia, with the
significant difference that its decays are not mediated through strong interaction but weak interaction due to its flavour quantum numbers  $B=-C=\pm 1$.
Theoretical investigations have been carried out on the properties of the $B_c$ meson, such as its lifetime, its decay constant, some of its form 
factors~\cite{Brambilla:2004wf}, 
based on OPE~\cite{Bigi:1995fs,Beneke:1996xe}, potential
models~\cite{Colangelo:1999zn,Kiselev:2000jc},
NRQCD and perturbative methods~\cite{Chang:1996jt,Lusignoli:1991bn,Brambilla:2000db,Brambilla:2004jw}, sum
rules~\cite{Kiselev:2003ds,Kiselev:2000nf,Kiselev:2000pp,Kiselev:2002vz}, 
or lattice gauge simulations~\cite{Davies:1996gi,Allison:2004hy,Shanahan:1999mv,Jones:1998ub}.
The properties of the $B_c$ meson will be
further scrutinized by the LHC experiments; the high luminosity of the LHC
machine opens the possibility to observe many $B_c$ decay channels beyond the
discovery one, in particular at LHCb.

This article is focused on the two-body non-leptonic charmless $B_c$
decays.  Indeed, the charmless $B_c$ decays with two light mesons ($\pi,
K^{(*)}, \eta^{(\prime)}, \rho, \omega, \phi$) in the final states come from
a single diagram: the initial $b$ and $c$ quarks annihilate into a charged
weak boson that decays into a pair formed of a $u$
and a $d/s$ quark, which hadronise into the two light mesons. This picture is
rather different from  processes such as $B_c\to J/\psi \pi$ for which the
initial $c$ quark behaves as a spectator. The recent  high-precision
measurements of the  $B_{u,d,s}$ and $D_{s}$ decays indicate that 
such annihilation processes can be significant, contrary to the theoretical expectation of its suppression in the heavy-quark limit. Indeed, fits of the data not taking into account annihilation processes are generally of poor quality. 

But the understanding of
these contributions remains limited. The theoretical computation of
annihilation diagrams is very difficult, so that the annihilation contributions are
often considered as a free parameter in these decays. On the other hand, in many $B_{u,d,s}$ decays, these annihilation
contributions come from several different operators (tree and penguin), and
they interfere with many different other (non-annihilation) diagrams, 
making it difficult to obtain an accurate value of annihilation by fitting experimental data. 
For this reason, the processes such as $B_d\to K^+K^-$, $B_d\to
D_s^-K^+, B_u\to D_s^-K^0$, where only the annihilation diagram contributes,
have been intensively worked out while the current experimental measurements are
still of limited accuracy.

The non-leptonic charmless $B_c$ decays which we discuss in this article can
have an important impact on this issue. We have 32 decay channels which come
from annihilation only, as mentioned above. Moreover, these decays involve a
single tree operator, which allows us to derive extremely simple relations
among the different decay channels. Finally, when  LHCb starts running and
observes the pattern of (or at least, provides bounds for) the branching
fractions  of these decays channels, it will certainly help us to  further
improve  our understanding of the annihilation contributions.

The remainder of the article is organized as follows. In section~\ref{sec:2},
we introduce the decay processes which we consider in this article.  In
section~\ref{sec:3}, we exploit $SU(3)$ flavour symmetry to derive relations
among amplitudes for non-leptonic charmless $B_c$ decays. In section~\ref{sec:4}, we discuss
theoretical issues related to this annihilation diagram and attempt to estimate its
size. In section~\ref{sec:5}, we discuss the prospects of searching 
non-leptonic charmless $B_c$ decays at LHCb and we conclude in
section~\ref{sec:6}. Two appendices are devoted to relating our work to
results from factorisation approaches.

%%%%%%%%%%%%%%%%%%%%%%%%%%%%%%%%%%
%
\section{{\large \bf Non-leptonic charmless $B_c$ decays as pure annihilation processes}}
\label{sec:2}
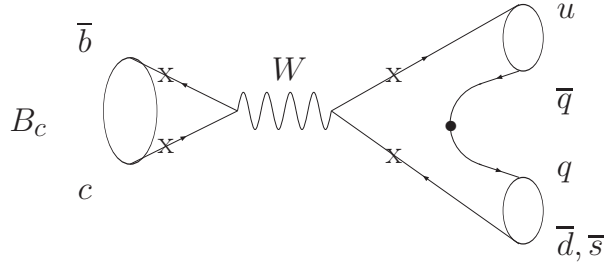
\begin{figure}[htbp]
\begin{center}
\scalebox{0.5}{
\begin{picture}(500,210)(0,-10)\setlength{\unitlength}{0.3mm}\Huge
\put(60,170){$\overline{b}$}\put(60,40){${c}$}\put(0,100){${B}_c$}
\put(480,200){$u$}\put(480,-10){$\overline{d},\overline{s}$}\put(480,60){$q$}\put(480,120){$\overline{q}$}
\put(230,145){$W$}%\put(510,160){$K^+$}\put(510,40){$\phi$}
\thicklines
\ArrowLine(170,100)(90,140)\ArrowLine(90,60)(170,100)
\Oval(90,100)(40,20)(0)
\Photon(170,100)(240,100){14}{4}
\ArrowLine(240,100)(380,180)\ArrowLine(380,0)(240,100)
\ArrowLine(380,130)(350,120)\ArrowLine(350,60)(380,50)
\put(382,100){\small\ding{108}}
\put(130,143){x}\put(130,80){x}
\put(330,143){x}\put(330,72){x}
\CArc(360,90)(31.5,105,255)
\Oval(383,155)(25,15)(0)\Oval(383,25)(25,15)(0)
\end{picture}}
\caption{Generic diagram for the non-leptonic charmless $B_c$ decays}
\label{fig:1}
\end{center}
\end{figure}

The diagram for the non-leptonic charmless $B_c$ decays is shown in fig.~\ref{fig:1} (the case of singlet states will be discussed below). 
The initial $\overline{b}$ and $c$ quarks annihilate into ${u}$ and $\overline{d}$ or $\overline{s}$ quarks, which form two light mesons by hadronising with a pair of $q\overline{q}$ ($q=u, d, s$) emitted from a gluon. 
There are 32 decay channels of this kind if we consider only the lightest pseudoscalar and vector mesons. In the case of two outgoing pseudoscalar mesons (PP), there are 4 modes with strangeness one: 
$$K^+ \pi^0, K^+ \eta, K^+ \eta^{\prime}, K^0 \pi^+$$ 
and 4 modes with strangeness zero: 
$$\pi^+ \pi^0, \pi^+ \eta, \pi^+ \eta^{\prime}, K^+ \bar{K}^0$$ 
The same applies for two vectors (VV) up to the obvious changes: 
$$K^{*+} \rho^0, K^{*+} \phi, K^{*+} \omega, K^{*0} \rho^+, 
\rho^+ \rho^0, \rho^+ \phi, \rho^+ \omega, K^{*+} \bar{K}^{*0}$$
In the case of VP decays, one can get two decay modes from one in PP
decays, depending on the pseudoscalar meson which is turned into a vector
one, yielding 8 strange decay modes ($\Delta S=1$ processes):
$$K^{*+} \pi^0, K^{*+} \eta, K^{*+} \eta^{\prime}, K^{*0} \pi^+,
\rho^0 K^+, \phi K^+, \omega K^+, \rho^+ K^0$$ 
and 8 non-strange decay modes ($\Delta S=0$ processes):
$$\rho^+ \pi^0, \rho^+ \eta, \rho^+ \eta^{\prime}, K^{*+} \bar{K}^0, 
\rho^0 \pi^+, \phi \pi^+, \omega \pi^+, \bar{K}^{*0} K^+$$ 
In the next section, we describe these decay channels in terms of a few
reduced amplitudes using $SU(3)$ flavour symmetry. Similar expressions have
been obtained for the charmless $B_{u, d}$ decays, which have been very useful to disentangle the rather complicated decay amplitudes of these decays containing many different contributions (tree, penguin, emission, annihilation, etc\ldots)~\cite{Chiang:2008vc, Chiang:2004nm,Chiang:2003pm}. Comparing to the case of $B_{u, d}$ decays, the $SU(3)$ relations for the $B_c$ decays are extremely simple, as it comes from only a single tree-annihilation diagram as mentioned earlier.

The theoretical computation of the process shown in Fig.~\ref{fig:1} amounts to determining the matrix element: 
\be
\langle h_1 h_2 | {\mathcal{H}}_{\rm eff} |B_c \rangle \label{eq:1}
\ee
where ${\mathcal{H}}_{\rm eff}$ is the effective Hamiltonian which we discuss later on. 
This matrix element contains contributions coming from the  $q\overline{q}$ state \emph{i)} produced perturbatively from one-gluon exchange linking the dot and one of the crosses in Fig.~\ref{fig:1}) and \emph{ii)} produced through strong interaction in the non-perturbative regime. 
Which type of these two contributions dominates this matrix element is an important issue in the theoretical computation of the hadronic $B$ decays.   In many approaches to non-leptonic $B$ decays~\cite{Keum:2000ph,Ciuchini:2001gv,Beneke:2003zv,Bauer:2004tj,Chay:2007ep}, 
it has been pointed out that annihilation diagrams may be sizable, with a
large imaginary part, so that they have  an important impact on the
phenomenology of CP violation in $B$ decays. Indeed, their contributions seem
to be needed to bring agreement between theoretical computations and
experimental results. There might be a significant difference in the
annihilation contributions for  $B$ and $B_c$ decays since the $B_c$ is likely 
to be considered as a heavy-heavy system rather than a heavy-light one. We will discuss the theoretical estimation of the annihilation diagram in more detail in section \ref{sec:4}. 

%%%%%%%%%%%%%%%%%%%%%%%%%%%%%%%%%%
%
\section{{\large \bf Relations from $SU(3)$ flavour symmetry}}
\label{sec:3}

In this section, we derive relations among the decay channels relying on the
$SU(3)$ flavour symmetry between $u$-, $d$- and
$s$-quarks. Following~\cite{Zeppenfeld:1980ex}, we first write down the
charmless $B_c$ decays in terms of the reduced amplitudes using the
Wigner-Eckart theorem. 

Let us first see the possible  $SU(3)$
representation of the external states. The initial state, $B_c^+$ is a singlet under $SU(3)$, whereas
the outgoing state is the product of two mesons, which can be either both in
the octet representation or in one octet and one singlet representations. We
have therefore outgoing states which transform as
\begin{equation}
8 \times 8 = 1 + 8_S + 8_A + 10 + 10^* \qquad
1 \times 8 = 8_I
\end{equation}
where the subscripts allow one to distinguish between the three different
octet representations involved. We sandwich the operators induced by the weak
interaction Hamiltonian between  these external states to obtain the
amplitudes for  the $B_c$ decays. The weak Hamiltonian for such transitions is
given by:
\be
{\mathcal{H}}_{\rm eff}=-\frac{G_F}{\sqrt{2}}\left[V_{ud}V_{cb}^*{\mathcal{O}}^{\Delta S=0}+V_{us}V_{cb}^*{\mathcal{O}}^{\Delta S=1}\right]
\ee
where the operators are: 
\bea
{\mathcal{O}}^{\Delta S=0}&=&\overline{u}\gamma_{\mu}(1-\gamma_5)d\ 
\overline{c}\gamma^{\mu}(1-\gamma_5)b \\
{\mathcal{O}}^{\Delta S=1}&=&\overline{u}\gamma_{\mu}(1-\gamma_5)s\ 
\overline{c}\gamma^{\mu}(1-\gamma_5)b 
\eea
These two operators are both $SU(3)$ octets and have the following $SU(3)$ tensor structures: 
\bea
{\mathcal{O}}^{\Delta S=0}:&\quad& (Y, I, I_3)=(0,1,1) \\
{\mathcal{O}}^{\Delta S=1}:&\quad& (Y, I, I_3)=(0,1/2,1/2) 
\eea 
where $(Y, I, I_3)$ denotes hypercharge, isospin and isospin projection respectively.
Since $B_c$ charmless decays involve only operators in an octet
representation, one can use the Wigner-Eckart theorem to express all the decay amplitudes in terms of 
three reduced matrix elements: 
\begin{itemize}
\item a reduced amplitude $S = \langle 8_S || {\mathcal{O}}^{8} || 1 \rangle $
from the symmetric product of the two incoming octet mesons. 
\item a reduced amplitude $A = \langle 8_A || {\mathcal{O}}^{8} || 1 \rangle $ 
from the antisymmetric product of the same representations
\item a reduced amplitude $I = \langle 8_I || {\mathcal{O}}^{8} || 1 \rangle $ 
from the product of an octet and a singlet meson.
\end{itemize}
The operator ${\mathcal{O}}$ can be ${\mathcal{O}}^{\Delta S=0}$ or
${\mathcal{O}}^{\Delta S=1}$. 
Note that the values of the reduced  quantities $S, A, I$ are in principle different for the $PP$, $VP$ or $VV$ final states. 
The Wigner-Eckart theorem requires one to compute the Clebsch-Gordan coefficients
describing the projection of a given 
$8\times 8$ and $8\times 1$ final state onto the two octet operators of
interest. These coefficients can be easily determined by combining
the usual $SU(2)$ Clebsch coefficients with the so-called isoscalar coefficients given in
ref.~\cite{deSwart:1963gc}. 

Finally, we must consider the different symmetry properties of the out-going
states ($P$ and $V$) as discussed in ref.~\cite{Gronau:1996ga}. For $PP$ decays, where the wave function of the final state is symmetric, 
only $S$ contributes,  apart from the case of final states containing $\eta$ or $\eta'$ where both $S$ and $I$ are present. 
For $VP$ decays, the amplitude gets contributions from 
$S$ and $A$ (and $I$ for final states containing $\eta,\eta',\omega$ or $\phi$).
For $VV$ decays, there are three amplitudes corresponding
to the three possible polarisations (or equivalently partial waves)
allowed for the outgoing state.
The wave function is symmetric for $S$ and $D$ waves and
antisymmetric for $P$ wave, so that the matrix element $S$ contributes to $S$ and $D$ waves, 
whereas $A$ contributes to $P$ waves. $I$ contributes only to $S$ and $D$ waves of outgoing states containing $\phi,\omega$ mesons.

A comment is in order on the mixing of the mesons containing $SU(3)$ singlet
states. The $\eta,\eta^{\prime},\omega,\phi$ mesons are mixtures of  the
$SU(3)$ octet ($\eta^8$ or $\omega^8$) and singlet ($\eta^0$ or $\omega^0$)
flavour states 
\bea
|\eta (\omega) \rangle &=& \cos\theta_{p (v)} |\eta^8 (\omega^8)\rangle +\sin\theta_{p (v)} |\eta^0 (\omega^0)\rangle \\
|\eta^{\prime} (\phi) \rangle &=& -\sin\theta_{p (v)} |\eta^8 (\omega^8)\rangle +\cos\theta_{p (v)} |\eta^0 (\omega^0)\rangle 
\eea
where $|\eta^8\rangle$ and $|\omega^8\rangle$ have the flavour composition
${|u\bar{u}+d\bar{d}-2s\bar{s}\rangle}/{\sqrt{6}}$,
and $|\eta^0\rangle$ and $|\omega^0\rangle$ are ${|u\bar{u}+d\bar{d}+s\bar{s}\rangle}/{\sqrt{3}}$.
The determination of the mixing angles $\theta_{p, v}$ is an important
phenomenological issue in understanding the nature of these particles. Since
we do not aim at a high accuracy in our $SU(3)$ analysis, we will adopt the
following values for the mixing angles which are not very far from the phenomenological determinations:  
\be
\tan\theta_p=\frac{1}{2\sqrt{2}}, \qquad \tan \theta_v= \sqrt{2}. 
\ee
These angles correspond to the ideal mixing for the vector sector:
\be
\omega=(u\bar{u}+d\bar{d})/\sqrt{2} \qquad \phi=s\bar{s}
\ee 
and also yield a simple expression of the pseudoscalar mesons:
\be
\eta=(u\bar{u}+d\bar{d}-s\bar{s})/\sqrt{3} \qquad 
\eta^{\prime}=(u\bar{u}+d\bar{d}+2s\bar{s})/\sqrt{6}
\ee
The non-ideal mixing of the $(\eta,\eta^\prime)$ mesons is linked to the $U(1)_A$ anomaly (see, e.g., refs~\cite{Feldmann:1998vh,Feldmann:1999uf}).
This value of the pseudoscalar mixing angle $\theta_p=\arctan (2\sqrt{2})^{-1}\simeq
-19.5^{\circ}$ is close to phenomenological determination,
e.g. from the $J/\psi$ radiative decays, $\theta_p\simeq
-22^{\circ}$~\cite{Gerard:2004gx}. Let us stress that for the light
mesons, we take the same phase conventions as in ref.~\cite{Beneke:2003zv},
so that some amplitudes have a minus sign with respect to those obtained
from ref.~\cite{deSwart:1963gc}~\footnote{In detail,
 $(-\bar{u},\bar{d},\bar{s})$ transform as an 
anti-triplet~\cite{Gronau:1996ga}, which means that there is a $(-1)$ phase
between the conventions of refs.~\cite{deSwart:1963gc}
and~\cite{Beneke:2003zv} for the pseudoscalar mesons 
$\pi^-,\pi^0,K^-,\eta,\eta^{\prime}$ and the vector mesons 
$\rho^-,\rho^0,K^{*-},\phi,\omega$. We have multiplied all the amplitudes by a
further $(-1)$ factor, so that the differences between our results and those
obtained using ref.~\cite{deSwart:1963gc} are limited to a $(-1)$ factor for 
the decay amplitudes for $K^0 \pi^+$, $K^+\bar{K}^0$ and 
their vector counterparts.}.

%%%%%%%%%%%%%%%%%%%%%%%%%%%%%%%%%%
\subsection{$PP$ modes}

Taking into account the Clebsch-Gordan coefficients together
with the issue of octet-singlet mixing, we obtain the following amplitudes
for the $PP$ modes, we have
$$
\begin{array}{ccccc}
{\rm Mode} & {\rm Amplitude}            & \qquad &  {\rm Mode} & {\rm Amplitude}\\
K^+\pi^0  &  \sqrt{\frac{3}{10}} S^{PP} && \pi^+ \pi^0 & 0 \\
K^0 \pi^+ &  \sqrt{\frac{3}{5}} S^{PP}  && K^+ \bar{K}^0 & \sqrt{\frac{3}{5}} S^{PP}\\
K^+ \eta  & -\frac{2}{3\sqrt{5}} S^{PP} + \frac{\sqrt{2}}{3} I^{PP} &&
 \pi^+ \eta  & \frac{4}{3\sqrt{5}} S^{PP} + \frac{\sqrt{2}}{3} I^{PP} \\
K^+ \eta^{\prime} & \frac{1}{3\sqrt{10}} S^{PP} + \frac{4}{3} I^{PP} &&
\pi^+ \eta^{\prime} & -\frac{1}{3}\sqrt{\frac{2}{5}} S^{PP} + \frac{4}{3} I^{PP}\\
\end{array}
$$
Here and in the following tables, these amplitudes must be multiplied by  $G_F/\sqrt{2}$ and also the appropriate CKM factor $V_{uD}V_{cb}^*$ with $D=d$ or $s$. We notice the relations
\begin{equation}
A(B_c^+\to K^0\pi^+)=\sqrt{2}A(B_c^+\to K^+\pi^0)=\hat\lambda A(B_c^+\to
K^+\bar{K}^0)
\end{equation}
with the Cabibbo-suppressing factor $\hat\lambda=V_{us}/V_{ud}$. The above relations are valid in the exact $SU(3)$ limit (for instance,
we have $S^{PP}=S^{K^+\pi^0}=S^{K^0\pi^+}=S^{K^+\overline{K}^0}$). Obviously, these relations have some interest only if the size of  $SU(3)$ breaking remains limited -- we will discuss this issue in Sec.~\ref{sec:4}. 

%%%%%%%%%%%%%%%%%%%%%%%%%%%%%%%%%%
\subsection{$VP$ modes}

For the $VP$ modes, we have for the strange modes
$$
\begin{array}{ccccc}
{\rm Mode} & {\rm Amplitude} & \qquad &  {\rm Mode} & {\rm Amplitude}\\
K^{+*}\pi^0  &  \frac{1}{2}\sqrt{\frac{3}{5}} S^{VP} +\frac{1}{2\sqrt{3}} A^{VP} &&
\omega K^+   & 
  -\frac{1}{2\sqrt{15}} S^{VP} - \frac{1}{2\sqrt{3}} A^{VP} + \sqrt\frac{2}{3} I^{VP} \\ 
\rho^0 K^+   &  \frac{1}{2}\sqrt{\frac{3}{5}} S^{VP} -\frac{1}{2\sqrt{3}} A^{VP} &&
\phi K^+   & 
  \frac{1}{\sqrt{30}} S^{VP} + \frac{1}{\sqrt{6}} A^{VP} + \frac{1}{\sqrt{3}} I^{VP} \\ 
K^{*+} \eta  & 
  -\frac{1}{3}\sqrt\frac{2}{5} S^{VP} + \frac{\sqrt{2}}{3} A^{VP} + \frac{1}{3} I^{VP} &&
\rho^{+}K^0  & \sqrt{\frac{3}{10}}S^{VP}-\frac{1}{\sqrt{6}}A^{VP}\\
K^{*+} \eta^{\prime} & 
  \frac{1}{6\sqrt{5}} S^{VP} - \frac{1}{6} A^{VP} + \frac{2\sqrt{2}}{3} I^{VP} &&
K^{*0}\pi^+  & \sqrt{\frac{3}{10}}S^{VP}+\frac{1}{\sqrt{6}}A^{VP} 
\end{array} 
$$
and for the non-strange modes
$$
\begin{array}{ccccc}
{\rm Mode} & {\rm Amplitude} & \qquad &  {\rm Mode} & {\rm Amplitude}\\
\rho^+\pi^0  &  \frac{1}{\sqrt{3}} A^{VP} &&
\omega \pi^+   & 
  \frac{1}{\sqrt{15}} S^{VP} + \sqrt\frac{2}{3} I^{VP} \\
\rho^0 \pi^+   &  -\frac{1}{\sqrt{3}} A^{VP} &&
\phi \pi^+   & 
  -\sqrt\frac{2}{15} S^{VP}  + \frac{1}{\sqrt{3}} I^{VP} \\
\rho^+ \eta  & 
  \frac{2}{3}\sqrt{\frac{2}{5}} S^{VP}  + \frac{1}{3} I^{VP} &&
K^{*+}\bar{K}^0  & \sqrt{\frac{3}{10}}S^{VP}-\frac{1}{\sqrt{6}}A^{VP}\\
\rho^+ \eta^{\prime} & 
  -\frac{1}{3\sqrt{5}} S^{VP} + \frac{2\sqrt{2}}{3} I^{VP} &&
\bar{K}^{*0}K^+  & \sqrt{\frac{3}{10}}S^{VP}+\frac{1}{\sqrt{6}}A^{VP}
\end{array}
$$
providing the simple relations 
\begin{eqnarray}
A(B_c^+\to K^{*0}\pi^+) &=& \sqrt{2}A(B_c^+\to K^{*+}\pi^0)
=\hat\lambda A(B_c^+\to \bar{K}^{*0} K^+)\\
A(B_c^+\to \rho^+ K^0) &=& \sqrt{2}A(B_c^+\to \rho^0 K^+) =
\hat\lambda A(B_c^+\to K^{*+} \bar{K}^0)
\end{eqnarray}

It should be noted that the amplitude $I^{VP}$ can be significantly different for the processes involving the vector singlet ($\phi, \omega$) and the pseudoscalar singlet ($\eta, \eta^{\prime}$) since it is known that the latter should receive a contribution from the anomaly diagram. This could induce a significant breaking of the above relations for channels involving  $\eta,\eta^\prime$.

%%%%%%%%%%%%%%%%%%%%%%%%%%%%%%%%%%
\subsection{$VV$ modes}

For the $VV$ modes, we have three different configurations for the outgoing mesons, labeled 
by their (common) helicity. The left-handedness of weak interactions and the fact that
QCD conserves helicity at high energies suggest that the longitudinal amplitude should dominate over
the transverse ones (corresponding to helicities equal to $\pm 1$).These helicity amplitudes can be combined linearly into $S$, $P$ and $D$ wave amplitudes, and in particular, the longitudinal amplitude
is a linear combination of only $S$ and $D$ waves.
$$
\begin{array}{ccc}
{\rm Mode} & S,D {\rm\ Amplitudes} & P {\rm\ Amplitude}\\
K^{*+}\rho^0 &  \sqrt{\frac{3}{10}} S^{VV}_{0,2} & \frac{1}{\sqrt{6}} A^{VV}_{1} \\
K^{*+} \omega  & 
  -\frac{1}{\sqrt{30}} S^{VV}_{0,2} + \frac{2}{\sqrt{3}} I^{VV}_{0,2}& -\frac{1}{\sqrt{6}}A^{VV}_{1} \\ 
K^{*+} \phi  & 
  \sqrt{\frac{1}{15}} S^{VV}_{0,2} + \sqrt{\frac{2}{3}} I^{VV}_{0,2}& \sqrt{\frac{2}{3}}
  A^{VV}_{1} \\ 
K^{*0} \rho^+ & 
  \sqrt{\frac{3}{5}} S^{VV}_{0,2} & \frac{1}{\sqrt{3}} A^{VV}_{1} \\ 
\rho^+\rho^0 & 0 & \sqrt{\frac{2}{3}} A^{VV}_{1} \\
\rho^+ \omega  & 
  \sqrt\frac{2}{15} S^{VV}_{0,2} + \frac{2}{\sqrt{3}} I^{VV}_{0,2}& 0 \\ 
\rho^+ \phi  & 
  -\frac{2}{\sqrt{15}} S^{VV}_{0,2} + \sqrt{\frac{2}{3}} I^{VV}_{0,2}& 0 \\ 
K^{*+} \bar{K}^{*0} & 
  \sqrt{\frac{3}{5}} S^{VV}_{0,2} & -\frac{1}{\sqrt{3}} A^{VV}_{1} 
\end{array} 
$$
where the subscript denote the partial wave under consideration $\ell=0,1,2$. In particular, we
have the interesting relations 
\begin{eqnarray}
A(B_c^+\to K^{*0}\rho^+) &=& \sqrt{2}A(B_c^+\to K^{*+}\rho^0)\\
\hat\lambda A(B_c^+\to K^{*+} \bar{K}^0) &=& \sqrt{2}(-1)^\ell A(B_c^+\to K^{*+} \rho^0)
\end{eqnarray}

%%%%%%%%%%%%%%%%%%%%%%%%%%%%%%%%%%
\subsection{Zweig rule}

In the above expressions, the normalization between $S, A$ and $I$ amplitudes
are different: the former is related to the $8\times 8$ representation and
the latter to $1\times 8$. One may relate these two amplitudes
by means of the Zweig rule for the $\Delta S=0$ processes
involving $\phi$. At the level of quark diagrams, one can see 
that the $B_c^+\to \phi\pi^+(\rho^+)$ process cannot come from the diagram in
fig.~\ref{fig:1} -- since $\phi$ is a pure
$s\bar{s}$ state. The only production process come from
non-planar diagrams
diagram similar to fig.~\ref{fig:1}, but with a different combination of
quarks into the outgoing mesons: the $u\bar{d}$ quarks produced from the $W$
go into $\pi^+(\rho^+)$ whereas the $\phi$ meson is made of a
$s\bar{s}$-pair  produced from vacuum. Such a non-planar diagram is expected to be
suppressed, especially for vector mesons such as $\phi$ (at least three
gluons are needed perturbatively, and it is $1/N_c$ suppressed in the
limit of a large number of colours). 

Assuming that the amplitudes for  $B_c^+\to \phi\pi^+$ and $B_c^+\to
\phi\rho^+$ vanish, one obtain the following relations:
\begin{equation}
I^{VP}=\sqrt{\frac{2}{5}} S^{VP} \qquad I^{VV}_{0,2}=\sqrt{\frac{2}{5}} S^{VV}_{0,2}
\end{equation}
providing simpler expressions for the following $VP$ decay amplitudes
$$
\begin{array}{ccccc}
{\rm Mode} & {\rm Amplitude} & \qquad & {\rm Mode} & {\rm Amplitude}\\
K^{*+} \eta  & 
  \frac{\sqrt{2}}{3} A^{VP} &&
\rho^+ \eta  & 
  \sqrt{\frac{2}{5}} S^{VP}  \\
K^{*+} \eta^{\prime} & 
  \frac{3}{2\sqrt{5}} S^{VP} - \frac{1}{6} A^{VP} &&
\rho^+ \eta^{\prime} & 
  \frac{1}{\sqrt{5}} S^{VP}  \\
\omega K^+   & 
  \frac{1}{2}\sqrt\frac{3}{5} S^{VP} - \frac{1}{2\sqrt{3}} A^{VP}&&
\omega \pi^+   & 
  \sqrt\frac{3}{5} S^{VP} \\
\phi K^+   & 
  \sqrt{\frac{3}{10}} S^{VP} + \frac{1}{\sqrt{6}} A^{VP} &&
\phi \pi^+   & 
  0
\end{array}
$$
so that
\begin{equation}
A(B_c^+ \to \rho^+\eta)=\sqrt{2} A(B_c^+ \to \rho^+\eta')
\end{equation}
Although we assume here that the $I^{VP}$ amplitude is the same for the processes involving the vector singlet ($\phi, \omega$) and the pseudoscalar singlet ($\eta, \eta^{\prime}$), as required by $SU(3)$ symmetry, this assumption is broken for the pseudoscalar singlets due to the anomaly (seen for instance in the mass difference between $\eta$ and  $\eta^{\prime}$). At the quark level,  
the detached diagram for the pseudoscalar singlet states
($\eta,\eta^{\prime}$) cannot be neglected  since there is the 
well-known anomaly contribution  modifying the previous relation:
\begin{equation}
A(B_c^+ \to \rho^+\eta)-\frac{1}{3}\Delta I^{VP}=\sqrt{2}\left[A(B_c^+ \to \rho^+\eta')-\frac{2\sqrt{2}}{3}\Delta I^{VP}\right]
\end{equation}
where $\Delta I^{VP}$ denotes a potentially large $1/N_c$-suppressed anomaly contribution.

Keeping the same caveat in mind, we can simplify some $VV$ amplitudes 
$$
\begin{array}{ccc}
{\rm Mode} & S,D {\rm\ Amplitudes} & P {\rm\ Amplitude}\\
K^{*+} \omega  & 
  \sqrt{\frac{3}{10}} S^{VV}_{0,2} & -\frac{1}{\sqrt{6}}A^{VV}_{1} \\ 
K^{*+} \phi  & 
  \sqrt{\frac{3}{5}} S^{VV}_{0,2} & \sqrt{\frac{2}{3}}  A^{VV}_{1} \\ 
\rho^+ \omega  & 
  \sqrt\frac{6}{5} S^{VV}_{0,2} & 0 \\ 
\rho^+ \phi  &  0 & 0 \\ 
\end{array} 
$$
with the obvious relations among partial waves.

%%%%%%%%%%%%%%%%%%%%%%%%%%%%%%%%%%%%%%%%%%%%%%%%%%% 
\section{{\large \bf Estimating the branching ratios}}
\label{sec:4}
As mentioned in section \ref{sec:2},  a precise estimate of the matrix element for the annihilation diagram is an important theoretical issue in $B$ physics. Although the domination of the one-gluon exchange diagram has been argued in various theoretical frameworks for $B_{u,d,s}$ decays~\cite{Beneke:2003zv, Keum:2000ph}, it has not been investigated whether one-gluon exchange or other (non-perturbative) contributions dominate in $B_c$ decays. In this section, we provide branching ratio estimates for the non-leptonic charmless $B_c$ decays in two ways, by using experimental data on pure annihilation $B$ decays and by relying on the one-gluon picture \`a la QCD factorisation.
The branching ratios can then be readily obtained by the usual formulae 
\be
Br(B_c\to h_1h_2)={\small \frac{\sqrt{[M_{B_c}^2-(m_1+m_2)^2][M_{B_c}^2-(m_1-m_2)^2]}}{\Gamma_{B_c}^{\rm tot}16\pi M_{B_c}^3}}|\langle h_1 h_2 | {\mathcal{H}}_{\rm eff} |B_c \rangle |^2 
\ee
\corrE{and the expression obtained in the section \ref{sec:3} in terms of the reduced amplitudes is related to the matrix element  through }
\be
|\langle h_1 h_2 | {\mathcal{H}}_{\rm eff} |B_c \rangle |=G_F/\sqrt{2}|V_{ud(s)}V_{cb}^*|\times |R(h_1,h_2)|
\ee
where the reduced amplitude $R(h_1,h_2)$ involves the amplitudes listed in
sec.~\ref{sec:3}, expressed in terms of  ($S, A, I$).

%%%%%%%%%%%%%%%%%%%%%%%%%%%%%%%%%%%%%%%%%%%%
\subsection{Estimate from $B_d$ annihilation process}

There are two pure annihilation processes observed in heavy-light $B$ decays~\cite{Aubert:2006fha,Abe:2006xs,Morello:2006pv}:
\bea
Br(B^0\to K^+K^-)&=& (0.15^{+0.11}_{-0.10})\times 10^{-6} \\
Br(B^0\to D_s^- K^+) &=&(3.9\pm 2.2)\times 10^{-5}  
\eea
Although the large experimental errors do not allow us to draw any firm conclusion, these  data seem to indicate that the annihilation contribution is not negligible. 
We may attempt to use these decay channels to very roughly estimate the size of the non-leptonic charmless $B_c$ decays. Since we are interested in charmless final states, let us compare the $B_c\to K^+\overline{K}^0$ and the $B^0\to K^+K^-$ processes. Assuming naive factorization between initial and final states, the final-state contribution cancels out when taking the ratio of the amplitudes. As a result, we find:
\be
\frac{Br(B_c\to K^+\overline{K}^0)}{Br(B^0\to K^+K^-)}\simeq \underbrace{\left(\frac{V_{cb}}{V_{ub}}\right)^2}_{\sim 100}\underbrace{\left(\frac{f_{B_c}}{f_B}\right)^2}_{\sim 4}\underbrace{\frac{\tau_{B_c}}{\tau_{B_d}}}_{\sim 0.3}\frac{1}{\xi^2} .
\ee
The factor $\xi$ represents the difference due to the fact that the $B^0\to K^+K^-$ process comes from a diagram similar to Fig. \ref{fig:1} {\it but} the $W$ boson propagates in the $t$-channel\footnote{Here we neglect the small penguin diagram contribution to the $B^0\to K^+K^-$ decay.}. In the one-gluon picture, $\xi=C_1/C_2\simeq 4$ whereas it might be smaller once non-perturbative effects are included. Indeed these effects would add new contributions to both decays, but the $B_c$ decay would be more affected relatively, since its Wilson coefficient in the one-gluon picture is smaller. This very naive argument leads to the relation between these two branching ratios: 
\be
Br(B_c\to K^+\overline{K}^0)\simeq Br(B^0\to K^+K^-) \times \frac{1.2\times 10^2}{\xi^2}
\gsim Br(B^0\to K^+K^-) \times 7.5
\ee
The estimate for $Br(B_c\to K^+\overline{K}^0)$ using this relation depends on
the result on $ Br(B^0\to K^+K^-)$, which should be improved in the near
future. Taking the current central value of $ Br(B^0\to K^+K^-)$, we find a lower limit of $Br(B_c\to K^+\overline{K}^0)$ at the order of $10^{-6}$.
Using this result, one can estimate the Wigner-Eckart reduced matrix elements
$S^{PP}$:
\be
S^{PP}\gsim 0.085 {\rm\ GeV}^3  \qquad [(PP)=(\overline{K}^{0}K^+)]\label{eq:19}
\ee
where we used the following CKM central values: $V_{ub}=0.0035, V_{cb}=0.041$~\cite{Charles:2004jd}. 

As mentioned before, there is no good reason to assume that $S^{PP}$, $S^{PV}$ and $S^{VV}$ should be related. Since we are only looking for order of magnitudes, we will assume as a dimensional estimate that  $|S^{PP}|\simeq \sqrt{2} |S^{PV}| \simeq |S_0^{VV}|$.  We emphasise that these relations have no strong theoretical supports and are just meant as a way to extract order of magnitudes for the branching ratios. 
In addition, we assume the Zweig rule to determine the singlet contributions $I$ and we neglect the antisymmetric contributions $A$, as well as transverse $VV$ amplitudes.
This provides the following branching ratios of interest, for instance:
\begin{eqnarray} 
[B_d\textrm{\ annihil}] 
&& BR(B_c\to\phi K^+)\simeq {\mathcal{O}}(10^{-7}-10^{-8}), \quad BR(B_c\to\bar{K}^{*0} K^+)\simeq {\mathcal{O}}(10^{-6}) \nonumber \\
&&  BR(B_c\to\bar{K}^{0} K^+)\simeq  {\mathcal{O}}(10^{-6}),  
\quad  BR(B_c\to\bar{K}^{*0} K^{*+})\simeq  {\mathcal{O}}(10^{-6})
\end{eqnarray}
The suppression of the $\phi K^+$ channel is due to the small CKM factor for
the $\Delta S=1$ processes, Cabibbo-suppressed compared to $\Delta S=0$.

\subsection{Estimate from one-gluon exchange model}

A second method consists in a model based on one-gluon exchange, in close relation
with the model proposed in QCD factorisation to estimate annihilation contributions for the decays
of heavy-light mesons.
In this method, described in more detail in App.~A, the matrix element in eq. (\ref{eq:1}) can be given as: 
\be
\langle h_1 h_2 | {\mathcal{H}}_{\rm eff} |B_c \rangle 
=i\frac{G_F}{\sqrt{2}}V_{cb}^* V_{ud(s)}N_{h_1h_2} b_2(h_1,h_2)
\ee
where
\be
N_{h_1h_2}=f_{B_c}f_{h_1}f_{h_2}
\qquad 
b_2(h_1,h_2)=\frac{C_F}{N_C^2}C_2 A_i^1(h_1h_2). 
\ee
The function $A_i^1(h_1h_2)$ is estimated as the convolution of the  kernel given by one-gluon exchange diagrams and the distribution amplitudes of the initial and final state mesons. \corrE{While} the detailed computation of this function can be found in Appendix~A, we would like to emphasize a few differences in $B_c$ decays comparing to the $B_{u,d}$ decays that we found; \emph{i)} 
the $B_c$ decays are much simpler, since the only operator contributing is $O_2$,
and there is only one
combination of CKM factors 
($V_{cb}^* V_{uD}$ where $D=d,s$ depending on the 
strangeness of the outgoing state), 
\emph{ii)} the long-distance divergences, which
prevents us from estimating the annihilation contribution in $B_{u,d}$ decays, do not appear in $B_c$ annihilation.

We can give the numerical results for the following channels: $B_c\to \phi
K^+, B_c\to K^{*0}K^+, B_c\to \overline{K}^0K^+, B_c\to
\overline{K}^{0*}K^{*+}$. We start with the function $b_2(h_1h_2)$ which turns
out to be quite $SU(3)$ invariant:
\bea
b_2(\phi K^+)=1.5, &\quad& b_2(K^{*0} K^+)=1.4, \\
b_2(\overline{K}^{0} K^+)=1.6, &\quad& b_2(\overline{K}^{0*} K^{*+})=1.0 \label{eq:25-2}
\eea
Notice that $b_2(VP)=b_2(PV)$, and thus $A^{VP}=0$. One can see that the
$SU(3)$ breaking is rather small as argued in Appendix~A, while the difference
between $PP$ and $VP(VV)$ modes can be as large as +13(+38)\%. We next list
the normalization factors:
\bea
N_{\phi K^+}=0.014{\rm \ GeV}^3,&\quad& N_{K^{*0} K^+}=0.014{\rm \ GeV}^3,\\  N_{\overline{K}^{0} K^+}=0.010{\rm \ GeV}^3,&\quad& N_{\overline{K}^{0*} K^{*+}}=0.019{\rm \ GeV}^3
\eea
For these particular decay channels (no exchange between $\pi$ and $K$), the $SU(3)$ breaking is also small while the difference between $PP$ and $VP(VV)$ modes can be as large as  -40(-90)\%.  
Our numerical value for $S,I,A$ amplitudes for the above processes are:
\begin{eqnarray}
S^{VP}=0.036{\rm\ GeV}^3, &\quad& 
[VP=\phi K^+,\overline{K}^{*0} K^+]\\ 
S^{PP}=0.021 {\rm\ GeV}^3, &\quad& [PP=\overline{K}^{0} K^+]\\
S_0^{VV}=0.025{\rm\ GeV}^3 &\quad& [VV=\overline{K}^{0*} K^{*+}]
\end{eqnarray}
where we have neglected transverse $VV$ amplitudes.
We present the relation between $N_{h_1h_2}$ and $b_2(h_1h_2)$
functions and the Wigner-Eckart reduced matrix elements $S,I,A$ in App.~B. 
Assuming the $SU(3)$ breaking effect is negligible in $b_2(h_1,h_2)$, this relation allows us to estimate all the other $S, I, A$ amplitudes with the values given in Eq. (\ref{eq:25-2}) and the known values of decay constants for each final state. 

We obtain finally the following values of branching ratios:
\begin{eqnarray} 
[\textrm {One-gluon}] 
&& BR(B_c\to\phi K^+)=5\times 10^{-9}, \ \ \  BR(B_c\to\bar{K}^{*0} K^+)=9.0\times 10^{-8}\\
&&  BR(B_c\to\bar{K}^{0} K^+)=6.3\times 10^{-8},  BR(B_c\to\bar{K}^{*0} K^{*+})=9.1\times 10^{-8}
\end{eqnarray}
The contributions to $\rho^0 \pi^+$, $\phi \pi^+$ and $\rho^+\phi$ vanish in
our approximations, which means that these power-suppressed decays must have
significantly smaller branching ratios than the above ones. 
We do not quote any error bars on these results on purpose: we can easily estimate the uncertainties coming from our hadronic inputs, but certainly not the systematics coming from the hypothesis underlying our estimate (one-gluon approximation, asymptotic distribution amplitudes, neglect of $1/m_b$ and $1/m_c$ suppressed corrections, neglect of soft residual momentum of the heavy quarks in the $B_c$ meson). 

\subsection{Comparison of the two methods}

Our estimates of the branching ratio for e.g. the $B_c\to K^+\overline{K}^0$ in the above two ways are not consistent. This is clearly because two methods are conceptually different:
\begin{itemize}
\item The method based on $B_d$ annihilation treats the charm quark as massless. It takes into account some of the non-perturbative long-distance effects expected to occur in $B_d$ and $B_c$ decays, but treats in a very naive way the relation between matrix elements of the operators $O_1$ and $O_2$. It relies also on extremely naive assumptions concerning the respective size of matrix elements for $PP$, $VP$ and $VV$ modes.
\item The method based a perturbative one-gluon exchange treats the charm quark as heavy. It assumes the dominance from a specific set of diagrams computed in a perturbative way, but it provides a consistent framework to perform the estimation. 
\end{itemize}
It is well known that both kinds of estimates yield rather different results. This is illustrated by the fact that the estimate of $B_d\to K^+K^-$ in the annihilation models of QCD factorisation~\cite{Beneke:2003zv} (around $10^{-8}$ with substantial uncertainties) and perturbative QCD~\cite{Chen:2000ih}  is one order of magnitude below the current experimental average. Therefore, it is not surprising that our two methods
yield branching ratios differing by a similar amount.
There are well-known cases where final-state interaction can increase significantly 
estimates based on factorisation, for instance $B\to K\chi_c$~\cite{Pham:2005ih,Beneke:2008pi} or $D_s^+\to \rho^0\pi^+$~\cite{Fajfer:2003ag}.
An observation of the non-leptonic charmless $B_c$ decays will certainly have an important key to clarify such a controversy as well as further theoretical issues in computational methods for the annihilation diagram.

%%%%%%%%%%%%%%%%%%%%%%%%%%%%%%%%%%%%%%%%%%%%%%%%%%% 
\section{{\large \bf Search prospect at LHCb}}
\label{sec:5}

The LHC pp collider with the center of mass energy of 14 TeV has a large cross section for the $b\overline{b}$ hadro-production, which can be followed by the production of not only $B^0$, $B^+$ and $B_s^0$ mesons but also other $b$ hadrons such as, $\Lambda_b$ and $B_c$. The subtraction of the production of other known $b$ hadrons~\cite{pdg}
\begin{equation}
b\overline{b}\to (B_d: B_u: B_s: \Lambda_b'{\rm s})\simeq (42.2\pm 0.9\% : 42.2\pm 0.9\% : 10.5\pm 0.9\%: 9.1\pm 1.5\%)
\end{equation}
 leads to  $b\overline{b}\to (B_c)$ to be less than one \%. The LHCb experiment~\cite{LHCb:detpaper}, which is dedicated to  $B$ physics analyses with its optimized trigger scheme, allows one to detect the $b$-decay modes into hadronic final states. 

The theoretical estimate of the $B_c$ cross section is still under scrutiny. For the dominant $gg$-fusion process, there are two possible mechanisms: $gg\to b\bar{b}$ followed by the fragmentation, or $\bar{b}\to B_c b\bar{c}$. It is found that the latter dominates in the low-$p_T$ region which corresponds to the LHCb coverage~\cite{Gouz:2002kk,Kolodziej:1995nv}. The ${\mathcal{O}}(\alpha_s^4)$ computation of the $\bar{b}\to B_c b\bar{c}$ process predicts $\sigma(pp\to B_c^+X)\simeq 0.3-0.8\ \mu b$~\cite{LHCbnote} where the error comes from the uncertainty on theoretical inputs such as the choice of the $\alpha_s$ scale and the $B_c$ distribution function. Additional systematics could come from higher-twist and radiative corrections. In the following, we follow the LHCb value for the cross section $\sigma(B_c) = 0.4\ \mu \rm{b}$ but it must be noted that this value may be affected by a large uncertainty.  

We can now estimate the expected sensitivity for a specific channel. First, let us discuss which channel has the best potential for the detection. The best trigger  and reconstruction efficiencies with a large signal over background ratio can be achieved by the charged $K$- and/or $\pi$-tags (and by avoiding low-$p_T$ neutral particles) at LHCb. Since the initial $B_c$ carries an electric charge, all two-body $PP$ final states  contain one neutral  particle. The same remark applies for the $VV$ channels when one considers the subsequent decays of the vector particles into pairs of pseudoscalars. In this respect, $PV$ channels such as $B_c^+\to \phi K^+,\ \overline{K}^{*0}K^+,\  \overline{K}^{^0}\pi^+,\ \rho^0K^+,\ \rho^0\pi^+,\ \phi \pi^+ $ are the best candidates using the vector meson decays, $\phi\to K^+K^-, \ \overline{K}^{*0}\to K^- \pi^+, \ \rho^0\to \pi^+\pi^-$, leading to three charged tracks. Among these subsequent decays, the small widths of  $\phi$ and $\overline{K}^{*0}$ make the reconstructing  particularly easy comparing to e.g. $\rho^0$. On the theoretical side, our  Zweig rule argument forbids the $B_c^+\to \phi \pi^+$, whereas the $B_c^+\to \rho^0\pi^+$ channel comes only from the $A$ (asymmetric) amplitude which is also subdominant. Finally, taking into account the fact that the $\Delta S=1$ channels are Cabibbo suppressed, we draw the conclusion that the $B_c^+\to \overline{K}^{*0}K^+$ channel might be the best candidate for the detection. 

Since the selection criteria and trigger efficiencies are different for each channel, detailed simulations are necessary  in order to estimate the expected sensitivity for different channels. For example, such a study has been done for $B_c\to J/\psi \pi^+$~\cite{LHCbnote,LHCbnote2}. From the expected branching ratio $Br(B_c\to J/\psi \pi^+)\simeq 1$ \%, it was deduced that over a thousand of events are expected after the one year run of LHCb. By scaling this observation to the processes of interest, we can very roughly estimate that an assumption of  $Br(B_c^+\to \overline{K}^{*0} K^+)=10^{-6}$ yields a few events per year at LHCb. The analysis of LHCb data will thus allow to set first experimental limits on the non-leptonic charmless $B_c$ decays, and give hints on annihilation mechanisms in these decays. 

%%%%%%%%%%%%%%%%%%%%%%%%%%%%%%%%%%%%%%%%%%%%%%%%%%% 
\section{\large \bf Conclusions}
\label{sec:6}

In this paper, we have discussed non-leptonic charmless $B_c$ decays  into two light pseudoscalar or vector mesons. It turns out that a single tree annihilation diagram is responsible for all 32 processes, providing an interesting testing ground for annihilation. After discussing general aspects of the charmless $B_c$ decays, we have shown that the very simple nature of these decays allows us to describe them in terms of a few reduced amplitudes by exploiting $SU(3)$ flavour symmetry to relate various $PP$, $PV$ and $VV$ modes.

In order to discuss a possible search for charmless non-leptonic $B_c$ decays at LHCb, we have proposed two different theoretical estimates of these reduced matrix elements, either by comparison with $B_d$ annihilation processes or by a perturbative model based on the exchange of one-gluon. The two models yield a rather wide range of branching ratio predictions, from $10^{-6}$ to $10^{-7}$. The LHCb experiment has the potential to observe some of the decay channels (such as $B_c\to \phi K^+, \overline{K}^{*0}K^+$) if the branching ratio is at the larger side of these estimates.

From the theoretical point of view, a better understanding of annihilation diagrams is particularly important. They are often assumed to play a significant role in decays of heavy-light mesons, but they occur jointly with other kinds of diagrams, making it difficult to assess precisely their size. Furthermore, for the theoretical estimates of $B_{u,d}$ annihilation diagram in the QCD factorisation, there is an additional uncertainty caused by the infrared divergence occurring in its computation. It is worth mentioning that we found that such a divergence does not occur in the case of the $B_c$ annihilation diagram suggesting that predictions from models \`a la QCD factorisation for the $B_c$ decays should be more precise, and thus easier to confirm or reject. 

On the other hand, it has been discussed that the annihilation diagrams may be enhanced by  long-distance effects such as final-state interactions. Although only limited models of such effects have been proposed either for $D$ or for $B$ decays (the former likely more affected than the latter by such enhancements)\cite{Fajfer:2003ag,Pham:2005ih,Beneke:2008pi}, the observation of an unexpectedly large branching ratio for the $B_c$ annihilation would call for a reassessment of such long-distance contributions. An observation of  charmless non-leptonic $B_c$ decays at LHCb will certainly provide substantial information on these models, in complement with the observation of other decays such as $B_d \to K^+ K^-$ or $B_s \to \pi^+ \pi^-$.

%%%%%%%%%%%%%%%%%%%%%%%%%%%%%%%%%%%%%%%%%%%%%%%%%%%
\section*{\bf \normalsize Acknowledgments}
We would like to thank Marie-H\'el\`ene Schune for discussion. Work supported in part by EU Contract No.   MRTN-CT-2006-035482, \lq\lq FLAVIAnet'' and
by the ANR contract  \lq\lq DIAM'' ANR-07-JCJC-0031. The work of E.K. was supported by the European Commission Marie Curie Incoming International Fellowships under the contract MIF1-CT-2006-027144 and by the ANR (contract "LFV-CPV-LHC" ANR-NT09-508531).

%%%%%%%%%%%%%%%%%%%%%%%%%%%%%%%%%%%%%%%%%%%%%%%%%%% 
\appendix

\section{Short-distance model for weak annihilation}

As highlighted in the introduction, weak annihilation plays a
 significant role in $B_{u,d,s}$ non-leptonic decays, but it is difficult to
 estimate it accurately. A model to estimate this contribution was provided
 in the framework of QCD factorisation~\cite{Beneke:2001ev,Beneke:2003zv}, relying on the following hypothesis :
\begin{itemize}
\item the diagrams are dominated by the exchange of a single gluon, whose
 off-shellness is typically of order $O(\sqrt{\Lambda m_b})$
\item hadronisation effects are taken into account through light-cone
 distribution amplitudes (generally taken in their asymptotic form)
\item soft components are neglected.
\end{itemize}
Being power-suppressed in the heavy-quark limit, the weak-annihilation
contributions to $B_{u,d,s}$ non-leptonic decays cannot be factorised in
short- and long-distance effects (in general). Their evaluation within this
rough model exhibits endpoint divergences, which signals the presence of long-
distance contributions not taken into account properly. The divergent
integrals were regularised on the basis of dimensional analysis, which induces
a significant uncertainty on the estimate of the annihilation contribution.

We can follow a similar method to estimate annihilation in the case of the $B_c$
decay. Concerning the $B_c$ meson, we work in the limit where both $b$ and $c$
quarks are heavy (keeping $m_c/m_b$ fixed) and we set the momentum of the 
valence quarks to $p_b^\mu=m_b v^\mu$ and $p_c^\mu=m_c v^\mu$, neglecting the
soft components of the heavy-quark momenta (and consistently setting
$M_{B_c}=m_c+m_b$). Since we neglect the soft components of $p_b$ and $p_c$,
the integration over the $B_c$ meson distribution amplitude is trivial and
yields $f_{B_c}$. The diagrams to compute are not very difficult
and correspond to a gluon emitted from the $b$ anti-quark or the $c$
quark from the $B_c$ meson and converted into a light quark-anti-quark pair.
Following ref.~\cite{Beneke:2003zv}, we find in the case where
$(M_1,M_2)=(P,P),(P,V),(V,V)$ :
\begin{eqnarray} \label{eq:A1i}
&&A_1^i(M_1M_2)=
 \pi\alpha_s \int dx\,dy\,\\ \nonumber
&&\quad  \Bigg\{\phi_{M_1}(y)\phi_{M_2}(x)
   \left[\frac{1}{y[(\bar{x}+y)z_b-\bar{x}y]}
        -\frac{1}{\bar{x}[(\bar{x}+y)z_c-\bar{x}y]}
\right]\\
&&\qquad   +r^{M_1}r^{M_2}\phi_{m_1}(y)\phi_{m_2}(x)
   \left[\frac{2(1-z_b)}
                {(\bar{x}+y)z_b-\bar{x}y}
        -\frac{2(1-z_c)}
                {(\bar{x}+y)z_c-\bar{x}y}
\right] \nonumber
\end{eqnarray}
If $(M_1,M_2)=(V,P)$, one has to change the sign of the second (twist-4) term
above. $\phi_M$ and $\phi_m$ are twist-2 and twist-3 two-particle distribution
amplitudes of the meson $M$, and $r^M$ is the normalisation of the twist-3
distribution amplitude. In the case of pseudoscalar mesons, we have:
\begin{equation}
r^\pi=\frac{2m_\pi^2}{m_b \times 2m_q}\qquad 
r^K=\frac{2m_K^2}{m_b (m_q+m_s)}
\end{equation}
responsible for the 
chiral enhancement of twist-4 contributions for pion and kaon
outgoing states. In the case of vector mesons, we have:
\begin{equation}
r^V=\frac{2m_V}{m_b}\frac{f_V^\perp}{f_V}
\end{equation}

$z_b$ and $z_c$ denote the relative size of the $b$ and $c$-quark masses:
\begin{equation}
z_b=\frac{m_b}{m_b + m_c} \quad
z_c=1-z_b=\frac{m_c}{m_b + m_c}
\end{equation}
Their appearance allows one to distinguish the diagram of origin
(corresponding to a gluon emitted from the $b$ or the $c$
quark line).

Eq.~(\ref{eq:A1i}) is in agreement with the expressions obtained in 
refs.~\cite{Beneke:2003zv} and~\cite{Beneke:2006hg} in the limit $z_b\to 1$
(and $z_c\to 0$). 
To simplify further the discussion, we take the asymptotic expression for the
distribution amplitudes
\begin{equation}
\phi_P(x)=6x(1-x) \qquad
\phi_V(x)=6x(1-x) \qquad
\phi_p(x)=1 \qquad
\phi_v(x)=3(2x-1) \qquad
\end{equation}
The structure of the singularities in the kernel is 
due to the propagator of the gluon and seems quite complicated. But if we take
as an example
\begin{equation}
\int_0^1 d\bar{x}\, dy \frac{1}{(\bar{x}+y)z-\bar{x}y}
 =\int d\bar{x}\, 
  \frac{1}{z-\bar{x}}
      \log\left|\frac{z-\bar{x}+\bar{x}z}{\bar{x}z}\right|
\end{equation}
The function to be integrated is continuous at $x=z$ and has integrable
singularities for $x=0$ and $x=z/(1-z)$. The integration can therefore be
performed without problem as long as $z$ is different from $0,1/2$ and 1 (no
coalescence of singularities). For the $B_c$ meson, it
means that the twist-2 and the twist-4 contributions have 
no endpoint singularities and yield finite integrals. Therefore, there is no need to introduce models to regularise the divergent integrals like in the case of heavy-light mesons.

The corresponding expressions for the four cases are slightly tedious, but
they can be approximated to a very good accuracy through 
low-order polynomials in $\delta$, where
$z_b=0.76+\delta$ and $z_c=0.24-\delta$ 
(corresponding to $m_b=4.2$ and $m_c=1.3$ for $\delta=0$):
\begin{eqnarray}
&&A_1^i(PP)= \pi\alpha_s 
[(-22.83+4.84\delta+808.3 \delta^2+2507 \delta^3 + 3425 \delta^4)
\\ \nonumber
&&\quad  +r^{M_1}r^{M_2}
 (-8.23-3.65\delta+73.6 \delta^2 - 16.1 \delta^3 + 3575 \delta^4 + 16007\delta^5)]\\
&&A_1^i(PV)= \pi\alpha_s 
[(-22.83+4.84\delta+808.3 \delta^2+2507 \delta^3 + 3425 \delta^4)\\ \nonumber
&&\quad
 +r^{M_1}r^{M_2}
 (-19.15-123.5\delta-130 \delta^2 + 58.7 \delta^3+ 7982 \delta^4 + 39778\delta^5)]\\
&&A_1^i(VV)= \pi\alpha_s 
[(-22.83+4.84\delta+808.3 \delta^2+2507 \delta^3 + 3425 \delta^4)\\ \nonumber
&&\quad
 +r^{M_1}r^{M_2}
 (2.44+222.4\delta+1565 \delta^2 + 3386 \delta^3+ 5824 \delta^4 -
 80831\delta^5-421927 \delta^6)]
\end{eqnarray}
Within the set of approximations performed here, $A_1^i(PV)$ and $A_1^i(PV)$
are identical.

We use the above formulae to estimate a few branching ratios. We take our
inputs for the vector decay constants and the Wilson coefficient
$C_2(\sqrt{m_b \Lambda_h})=-0.288$ from ref.~\cite{Ball:2006eu}, and the rest of 
our inputs from ref.~\cite{Beneke:2003zv}. We take the value of the $B_c$
meson decay constant $f_{B_c}=395$ MeV taking the central value from 
ref.~\cite{Brambilla:2004wf}.

Let us comment on the $SU(3) $ breaking, which can be included in this QCD computation. We do not have the $SU(3)$-breaking effects coming from the distribution amplitude: for instance, a small $m_s$ correction makes the $K$ and $K^{(*)}$ distribution amplitudes slightly asymmetric. On the other hand, we have the breaking effect in the chiral enhancement parameter $r^M$. The $SU(3)$ breaking (e.g. comparison of $r^\pi, r^K$ or $r^\rho, r^{K^*}, r^\phi$) turns out to be relatively small. 
Another $SU(3)$ breaking arises from the following decay constants in the normalization factor $N_{h_1h_2}$~\cite{Amsler:2008zzb,Ball:2006eu}: 
\bea
%&f_\pi = 131 {\rm MeV}, \quad f_K =160 {\rm MeV}& \\
%&f_\rho = 209 {\rm MeV}, \quad f_{K^*} =218 {\rm MeV}, \quad f_{\omega} =187 {\rm MeV}, \quad f_{\phi} =221 {\rm MeV}.&
&f_\pi = (130.4\pm 0.2) {\rm MeV}, \quad f_K =(155.5\pm 0.8) {\rm MeV}& \\
&f_\rho = (216\pm 3) {\rm MeV},\  f_{K^*} =(220\pm 5) {\rm MeV}, \ f_{\omega} =(187\pm 5) {\rm MeV},\  f_{\phi} =(215\pm 5) {\rm MeV}.& \nonumber
\eea
The $SU(3)$ breaking in the decay constants for the vector mesons is rather small while there is a 24\% difference in $\pi$ and $K$ decay constants.

%%%%%%%%%%%%%%%%%%%%%%%%%%%%%%%%%%%%%%%%%%%%%%%%%%%

\section{Identification with results from QCD factorisation}

The expressions for all the decay channels considered in Sec.~\ref{sec:3} can be recovered from refs.~\cite{Beneke:2003zv},~\cite{Beneke:2006hg} and \cite{Bartsch:2008ps} if we
identify between the Wigner-Eckart reduced matrix elements $S,I,A$ and the $O_2$ reduced coefficients $b_2$. In these references, one must take the expressions for the decay amplitudes of $B_u$ decays
into the relevant final state, and pick up the $O_2$ contribution, which is the only remaining one
once the related $B_c$ decay is considered. If we perform this identification, we obtain:
\begin{eqnarray}
S^{PP}&=&\sqrt\frac{5}{3}N_{PP} b_2(PP)\\
I^{PP}&=&\sqrt\frac{2}{3}N_{PP} b_2(PP)\\
S^{VP}&=&\sqrt\frac{5}{6}N_{VP} (b_2(PV)+b_2(VP))\\
A^{VP}&=&\sqrt\frac{3}{2}N_{VP} (b_2(PV)-b_2(VP))\\
I^{VP}&=&\sqrt\frac{1}{3}N_{VP} (b_2(PV)+b_2(VP))\\
S_{S,D}^{VV}&=&\sqrt\frac{5}{3}N_{VV} b_2^{S,D}(VV)\\
A_P^{VV}&=&0\\
I_{S,D}^{VV}&=&\sqrt\frac{2}{3}N_{VV} b_2^{S,D}(VV)
\end{eqnarray}

%\thebibliography

\end{document}